\documentclass[fleqn,10pt]{wlscirep}
\usepackage[utf8]{inputenc}
\usepackage[T1]{fontenc}
\usepackage{amssymb}
\usepackage{amsmath}
\usepackage{hyperref}
\usepackage{latexsym}
\usepackage{mathtools}
\usepackage{mhchem}
\usepackage{relsize}
\usepackage{xcolor}
\title{Effect of Contact Inhibition Locomotion on confined cellular organization}

\author[1,*]{Harshal Potdar}
\author[2,3]{Ignacio Pagonabarraga}
\author[1,*]{Sudipto Muhuri}

\affil[1]{Savitribai Phule Pune University, Department of Physics, Pune, 411007, India}
\affil[2]{Departament de F\'{\i}sica de la Mat\`eria Condensada, Universitat de Barcelona, Mart\'{\i} i Franqu\`es 1, E08028 Barcelona, Spain}
\affil[3]{UBICS University of Barcelona Institute of Complex Systems,  Mart\'{\i} i Franqu\`es 1, E08028 Barcelona, Spain}
\affil[*]{sudipto@physics.unipune.ac.in}

\begin{abstract}
Experiments performed using micro-patterned one dimensional collision assays have allowed a precise quantitative analysis
of the collective manifestation of contact inhibition locomotion (CIL) wherein, individual migrating cells reorient their direction
of motion when they come in contact with other cells. Inspired by these experiments, we present a discrete, minimal 1D {\it Active} spin model that mimics the CIL interaction between cells in one dimensional channels. We analyze the emergent
collective behaviour of migrating cells in such confined geometries, as well as the sensitivity of the emergent patterns to driving
forces that couple to cell motion. In the absence of vacancies, akin to dense cell packing, the translation dynamics is arrested
and the model reduces to an equilibrium spin model which can be solved exactly. In the presence of vacancies, the interplay of
activity-driven translation, cell polarity switching, and CIL results in an exponential steady cluster size distribution. We define
a dimensionless Péclet number $Q$ - the ratio of the translation rate and directional switching rate of particles in the absence
of CIL. While the average cluster size increases monotonically as a function of $Q$, it exhibits a non-monotonic dependence
on CIL strength, when the $Q$ is sufficiently high. In the high $Q$ limit, an analytical form of average cluster size can be obtained
approximately by effectively mapping the system to an equivalent equilibrium process involving  clusters of different sizes
wherein the cluster size distribution is obtained by minimizing an effective Helmholtz free energy for the system. The resultant
prediction of exponential dependence on CIL strength of the average cluster size and $Q^{1/2}$ dependence of the average cluster
size is borne out to reasonable accuracy as long as the CIL strength is not very large.
\end{abstract}

\begin{document}

\flushbottom
\maketitle
\thispagestyle{empty}

\section*{Introduction}

A distinctive feature of  cell colonies is the emergence of collective organization, sensitive to the self propulsion characteristics of the individual constituent cells \cite{ignaref2,levineref14, igna-pnas,levine-pnas}. In general, the functionality of the tissues is determined by the ability of cells to self-organize  into a diverse spectrum of spatio-temporal arrangement and patterns \cite{FEBS, levine-pnas, igna-pnas, levine-prl}. Collective cell migration is driven by the interplay of self-propulsion characteristics and the cell-cell interactions \cite{igna-pnas}. Drastic transformation of the underlying arrangement of the cells is associated with cell morphogenesis and cancer progression \cite{igna-pnas}.
  
For many cell types it has been observed that when individual cells come in contact with other cell, their direction of self propulsion tends to get re-polarized away from the neighbouring cells \cite{cil1,cil2}. This process is also referred to as Contact Inhibition Locomotion (CIL) \cite{cil1, cil2, igna-pnas, levine-pnas}. The CIL interaction arises due to the fact that when two cells collide, the cell front adheres to the colliding cell, which obstructs further cell movement \cite{cil1,cil2}. This leads to repolarization of the cell's cytoskeleton. This in turn creates a new front away from the adhesion zone and facilitates the two cells to separate \cite{cil1,cil2,cadherin}. This inherent ability of the individual cell to reorient its direction of movement upon contact with other cells is an important determinant for multitude of cellular processes ranging from morphogenesis, tissue organization, wound healing and cancer progression among others \cite{igna-pnas,levine-pnas,ignaref2}. The interplay of the distinct interactions at play at the level of individual cells including CIL  provides the basis of the controlling mechanism which allows colonies of cells to organize themselves into diverse array of morphologies and functional characteristics \cite{igna-pnas}.  

Phenomenological models for the collective behaviour of cell colonies and tissue reorganization have built on active matter approaches \cite{viscek, schnitzer, soto, active-rev, active-prl,ignasoft, raghu}. In this context, agent-based models have proven to be particularly useful for describing the collective behaviour of broad class of actively driven systems ranging from bacterial suspension to cell colonies  and helped in providing crucial insights into the specific phenomenology of active matter \cite{agent1,agent2, igna-pnas, levine-pnas, levine-prl, soto}. On the other hand, active hydrodynamic models have served to highlight the unifying principles of active organization based on symmetry principles and conservation laws \cite{sriram-rev, hydro1, hydro2, ananyo}. These approaches to understand actively driven systems have been complemented by active vertex models \cite{vertex1, vertex2} and phase models \cite{phasefield} for some class of active systems. In particular, agent models which have explicitly incorporated the CIL interaction within their ambit, have been able to qualitatively explain tissue phenotypes \cite{igna-pnas, levine-pnas} and theoretically predict transitions between different phases of collective organization in the cell colonies \cite{igna-pnas}. While many of these studies have had reasonable success in providing a qualitative understanding of several types of cell transitions, exploration of full phase parametric space  for different cell types has remained an important open problem, owing to increased complexity arising from more detailed specification of cell-cell interaction rules. In this context, it may be noted that, while diverse agent based continuum models have been studied extensively, discrete driven lattice gas modeling approaches  \cite {driven1,driven2,driven3,active-prl} have
remained relatively unexplored in the context of cell colonies and tissues. It is worthwhile to point out the potential of such minimalist modeling approach. In particular, variants of discrete driven lattice gas models have been used to describe  transport across bio-membranes \cite{chou}, bidirectional cellular cargo transport \cite{sm-pre1, sm-pre2}, collective transport of motor proteins on biofilaments \cite{frey}, and growth process of fungal mycelium \cite{fungi1, fungi2, fungi3}. 

We also adopt a similar modeling approach to gain insight on the role of CIL in determining the organization of cells in quasi 1D settings, motivated by experiments performed with cells on 1D collision assays that have been designed using micropatterning techniques \cite{cil1,cil2,ananyo}. The restricted nature of cellular movement in such  assays allows for more precise identification of collision event of cells apart from enhancing the efficiency of CIL response since the collisions between the cells is head on \cite{cil1,cil2}. In particular, this experimental technique has been used to study and quantify the effect of CIL in cultured {\it Xenopus} Neural crest cells which are confined to move on micro-patterned fibronectin lines with very narrow width. The narrow width of these assays forces the cells to move along the narrow fibronectin line and to undergo a repolarization by $180^{\circ}$ due to CIL interaction \cite{ananyo}.

We put forward and study an {\it Active} spin model, belonging to the class of driven lattice gas models, which mimics the movement of individual cell and binary cell-cell interaction mediated by CIL in 1D assay. We use this model to investigate the dynamics of collective organization of cells and the clustering characteristics of cells that are subject to CIL interactions in such geometries.

\begin{figure}[h!]
	\centering
	\includegraphics[width = \linewidth, height = 4 cm]{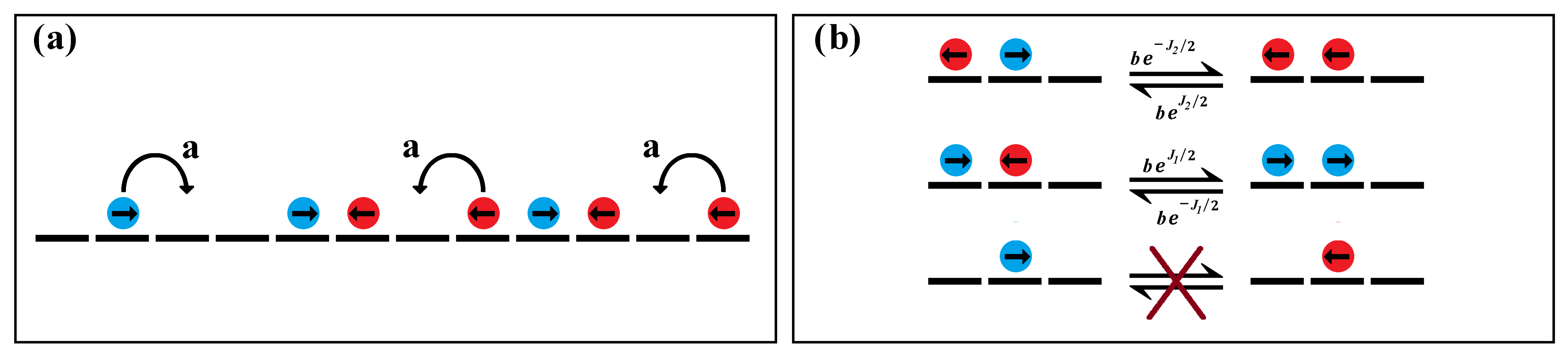}
	\caption{Schematic representation of the relevant dynamical process: (a) Translation process. (b) Polarity switching of a particle at the boundary of a cluster. A single particle bounded by vacancies does not switch its polarity.}
	\label{fig1}
\end{figure}
\section{Model and Methods}

We consider a discrete 1D lattice consisting of $L$  sites and represent individual cells as particles. The individual lattice sites can either be empty or be occupied by one particle. Each particle possesses discrete states of the individual polarization vector $\vec P$, associated with their direction of movement on the lattice. The polarization state of the particle at site $i$ maybe described in terms of a variable  $ \sigma_i$ which can take two different values, $\pm 1$, depending on whether  the polarization vector  points towards right or left direction on the lattice, respectively.  A general configuration of the system at a given time $t$ maybe represented as,

\begin{equation}
\rightarrow ~ \leftarrow~\rightarrow ~ \rightarrow~0 \rightarrow ~0 ~0\rightarrow ~ \leftarrow~0~\leftarrow ~ \leftarrow \nonumber
\end{equation}

Here, $(\rightarrow)$ corresponds to a particle moving to the right, while$(\leftarrow)$ corresponds to a particle moving to the left, and $0$ corresponds to an empty lattice site.

The primary characteristic of CIL is the propensity of the cells to align the direction of movement away from other neighbour cells. To mimic the effect of CIL in 1D channel, we consider a  short range interaction between the particles described by a nearest neighbour interaction potential, 

\begin{equation}
H = \sum_{i} J_1 \Theta( \sigma_{i} - \sigma_{i+1} ) -J_2 \Theta( \sigma_{i+1} - \sigma_{i} ), 
\label{eqn-H}
\end{equation}

where, $\Theta$ stands for the Heaviside function.
 
For this choice of $H$, the configurations of a pair of neighbouring particles on the lattice would have the following energy: 
\begin{eqnarray}
\rightarrow ~ \leftarrow ~~~~~~&\equiv& ~~~~~~E = + J_1  ~~(J_1 > 0) \nonumber\\
\leftarrow ~ \rightarrow ~~~~~~&\equiv& ~~~~~~E = -J_2  ~~(J_2 > 0)\nonumber\\
\rightarrow ~ \rightarrow ~~~~~~&\equiv& ~~~~~~E = 0  \nonumber\\
\leftarrow ~ \leftarrow ~~~~~~&\equiv& ~~~~~~E = 0\nonumber
\end{eqnarray}

where we have set $k_B T = 1$. Thus particle configurations for which polarization vectors of neighbouring particles face each other are disfavoured, while configurations for which polarization vectors are oppositely aligned are favoured. We define a  cluster as a continuous array of at least two particles that are bounded by vacancies at both ends. We consider that the switching dynamics of the particles between the different polarization states occurs only for clusters comprising of two or more particles, and is dictated by the interaction potential given by Eq. \ref{eqn-H}, such that the switching rates of particle's polarity, ${k_s\propto\exp(-\Delta E)}$, where $\Delta E$ is the energy difference between the two configurations. Thus the switching dynamics of the polarity state of a particle in the bulk of a cluster ( a particle bounded by other particles at its both end) maybe represented by, 

\begin{eqnarray}
\mathrm{\rightarrow~ \rightarrow~  \rightarrow } ~&\xrightleftharpoons[b e^{ \Delta J/2}]{b e^{ - \Delta J /2}}&~\mathrm{\rightarrow~\leftarrow ~ \rightarrow }\nonumber\\
\mathrm{\leftarrow~\leftarrow~  \leftarrow } ~&\xrightleftharpoons[b e^{\Delta J/2}]{b e^{-\Delta J /2}}&~\mathrm{\leftarrow ~ \rightarrow~\leftarrow }\nonumber\\
\mathrm{\rightarrow~\rightarrow~  \leftarrow } ~&\xrightleftharpoons[b]{b}&~\mathrm{\rightarrow~\leftarrow ~ \leftarrow }\nonumber\\
\mathrm{\leftarrow~  \leftarrow~\rightarrow } ~&\xrightleftharpoons[b]{b}&~\mathrm{\leftarrow~\rightarrow ~ \rightarrow } \nonumber
\end{eqnarray}

Here, $\Delta J = J_1 - J_2$ and $b$ is the switching rate of particle in a cluster between polarization state which have no energy difference.

Similarly, the switching dynamics of the polarity of the particle at the cluster boundary maybe represented by,

\begin{eqnarray}
\mathrm{\rightarrow~  \rightarrow } ~&\xrightleftharpoons[b  e^{-J_{2}/2}]{b e^{ J_{2}/2}}&~\mathrm{\leftarrow ~ \rightarrow }\nonumber\\
\mathrm{\rightarrow~  \rightarrow } ~&\xrightleftharpoons[b e^{ J_{1}/2}]{b e^{-J_{1}/2}}&~\mathrm{\rightarrow ~ \leftarrow }\nonumber\\
\mathrm{\leftarrow~  \leftarrow } ~&\xrightleftharpoons[b e^{-J_{2}/2}]{b e^{ J_{2}/2}}&~\mathrm{\leftarrow ~ \rightarrow }\nonumber\\
\mathrm{\leftarrow~  \leftarrow } ~&\xrightleftharpoons[be^{J_{1}/2}]{b e^{-J_{1}/2}}&~\mathrm{\rightarrow ~ \leftarrow } \nonumber
\end{eqnarray}

As far as the translation dynamics is concerned, the particles at site $i$ hops to the adjacent site  in the direction of its polarization vector, with a rate $a$, provided that site is vacant, i.e., if $\sigma_i = +1$ then the particle hops to site $i+1$ with rate $a$, if it is vacant. On the other hand if $\sigma_i = -1$, then it hops to site $i -1$ with rate $a$, if it is vacant.  These rules of particle movements maybe summarized as, 
 
\begin{eqnarray}
\rightarrow ~0~&\Longrightarrow &~0~\rightarrow\nonumber\\
0 ~ \leftarrow ~&\Longrightarrow &~\leftarrow~0\nonumber
\end{eqnarray}

\subsection{Connection and mapping to other models}

Its worthwhile to point out that in the absence of CIL interaction within clusters, which corresponds to setting $J_1$ and $J_2$ to zero, the model is very similar to Persistent Exclusion Process (PEP) \cite{soto,sm-pre2} with the crucial difference that while for our case a single particle ( which is not part of a cluster) does not undergo switching of their polarities, PEP allows for switching of a single particle in the lattice. 

The Hamiltonian introduced in Eq.\ref{eqn-H} can also be decomposed into an {\it Ising} like alignment term and an interaction term which mimics the effective repulsion interaction arising due to reorientation of the particle polarities. Specifically, 

\begin{equation}
H  = \mathlarger{\mathlarger{\sum}}_{i} - \left(\frac{J_1 - J_2}{4}\right) {\bf\sigma}_{i}\cdot{\bf\sigma}_{i+1} - \left(\frac{J_1 + J_2}{4}\right) ({\bf\sigma}_{i} - {\bf\sigma}_{i+1})\cdot {\bf \hat n}_{i,i+1},
\end{equation}

where, ${\bf \hat n}_{i,i+1}$ is an unit vector along ${\bf r_i}-{\bf r_{i+1}}$, corresponding to the position vector of the $i^{th}$ and $(i+1)^{th}$ site in the lattice. 

This is indeed similar to the interaction potential invoked in hydrodynamic modeling of CIL phenomenology \cite{ananyo}. The second term of this interaction potential is also similar to interaction potential arising out of coupling between polar order and nematic splay in nematic liquid crystal \cite{nematic}. For the symmetric case, $J_1 = J_2$, i.e., $\Delta J = 0$, the contribution of Ising-like alignment term vanishes and the interaction term associated with repulsion due to CIL essentially does not contribute to the energy in the bulk of the  system and it features only as a boundary term . As a consequence, for a fully packed lattice comprised of particles with such interaction, the specific heat of the system would be zero at all temperatures and the particle-particle correlation length, $\xi = 0$. Of course, in the presence of vacancies, which allows for active transport of particles on the lattice, this interaction term plays a vital role in determining the nature of organization on the lattice. 

For the asymmetric case, $J = J_1 = -J_2$, the model reduces to a 1D Antiferromagnetic Potts model for $J >0$ and to a 1D ferromagnetic Potts model  for $J < 0$. However, the choice of $J < 0$ for the model Hamiltonian does not correspond to physically realistic scenarios for CIL. CIL-motivated interactions require the signs of $J_1$ and $J_2$ to be positive in order to ensure that configurations where the polarization vectors of neighbouring cells point towards each other are disfavoured, while the opposite holds when the polarization vectors of neighbouring cells point away from each other. Such choice necessarily breaks the $Z_2$ symmetry of the underlying Hamiltonian, and consequently distinguishes itself from the underlying symmetry of the Potts Hamiltonian.

\subsection{Simulation Details}

We perform Monte Carlo (MC) simulations of the system of $N$ particles on $L$ lattice sites starting with random configuration of particles uniformly distributed on a 1D lattice with fixed number density $(\rho = N/L)$. For the initial configuration of particles on the lattice, we choose the polarization states of individual particles randomly with equal probability. We adopt a random sequential update procedure by choosing a site randomly with equal probability. For a site which is occupied by a particle, we perform the MC move for a particular process ( translation or directional switching) with the prescribed relative rates for the different processes. In order to ensure that the system settles to steady state, we wait for an initial transient time of $1000 \frac{L}{w}$ steps , where $w$ stands for the lowest rate ( among translation and switching). Thereafter we collect the statistics for the cluster sizes and other cluster characteristics, time averaging typically over at least $5000$ samples. These samples are collected with a time spacing of $10 \frac{L}{w}$ to ensure that the samples are uncorrelated.  

\section{Results}

\subsection{The fully packed state: A reduced equilibrium model}

\begin{figure*}[t]
	\centering
	\includegraphics[width = \linewidth, height = 7cm]{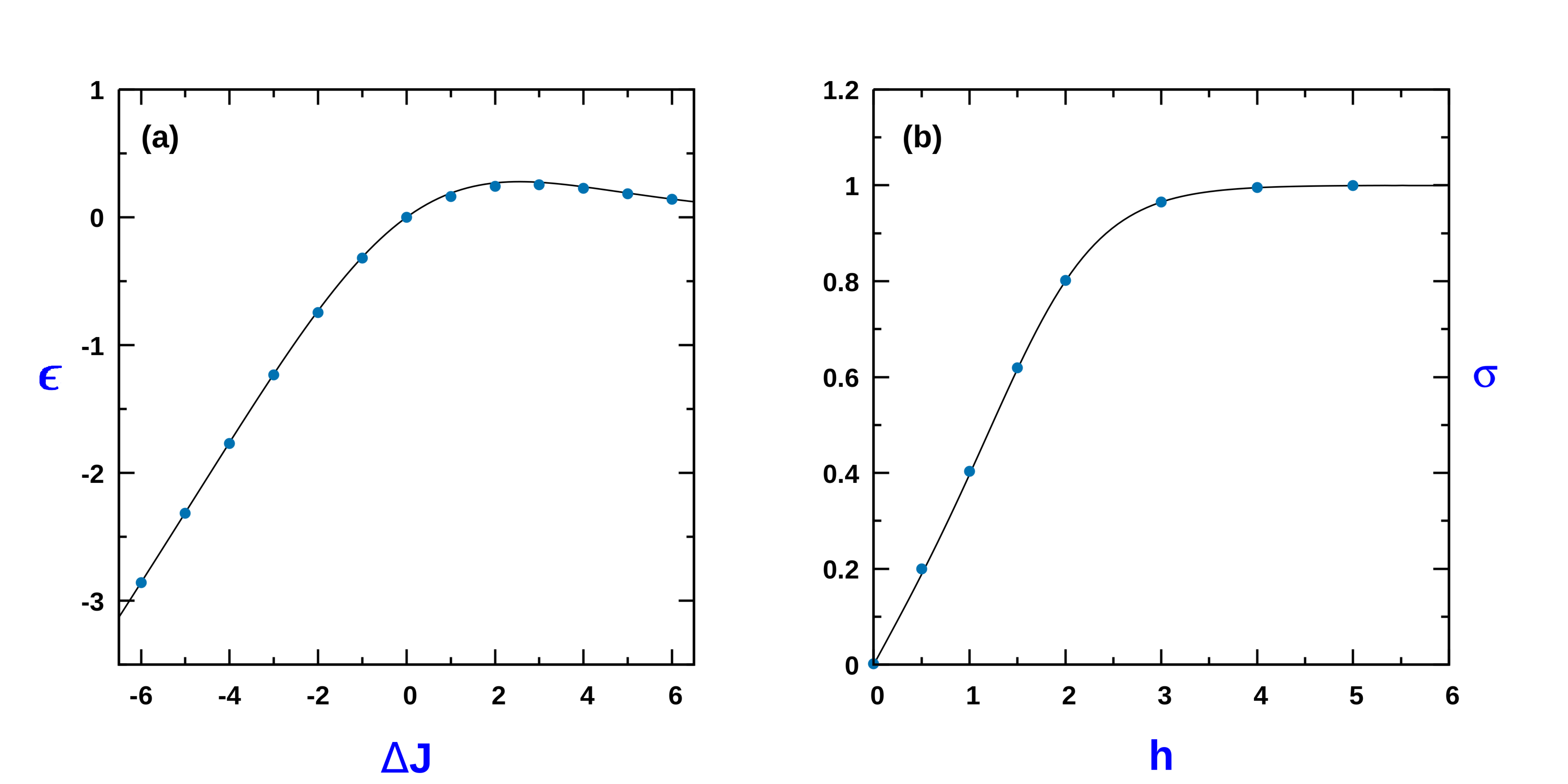}
	\caption{(a) Variation of the energy per particle, $\epsilon= E/N$  as a function of $\Delta J = J_1 - J_2$.  $\epsilon$ is expressed in units of $k_B T$, with $T$ being an effective temperature of the active system. (b) Variation of average polarization per particle, $\sigma$ as a function of external field $h$ for $\Delta J = 2$. Solid lines correspond to the analytical expression while the points corresponds to results obtained using MC simulations for $N=1000$.}
	\label{fig2}
\end{figure*}

In the experiments performed with MDCK cells on fibronectin coated strips, when all the adherent cells fill the strip corresponds to a confluent state \cite{ananyo}. In the confluent state, the dynamics at the boundaries  of the adherent cells can lead to motility \cite{ananyo}. However, from the perspective of our minimal model, where we treat the particles as entities with hardcore repulsion, in the absence of vacancies, the state of the system is such that the translation dynamics of the particles is arrested and the dynamics of the particles is restricted to switching process between different states of polarization of the particles. In this regime, the system behaves as an effective equilibrium system, whose properties are determined by the Boltzmann weight associated to the Hamiltonian of Eq. \ref{eqn-H} an effective temperature prescribed by the MC scheme. The corresponding form of the partition function, $Z_c$, maybe expressed as,

\begin{equation}
Z_c = \sum_{[\sigma_i]} \exp \left[ -\sum_{i} J_1 \Theta( \sigma_{i} - \sigma_{i+1} ) -J_2 \Theta( \sigma_{i+1} - \sigma_{i} )  \right], \nonumber
\end{equation}
\noindent where the corresponding transfer matrix for $Z_c$ reads,
\[
T=
  \begin{bmatrix}
    1 & e^{-J_{1}} \\\\
    ~e^{J_2} & 1 
  \end{bmatrix}
\]

Using standard techniques of Transfer Matrix method, in the thermodynamic limit of $N\rightarrow \infty$, we obtain 
\begin{equation}
Z_c = {\left[ 1 + \exp \left( \frac{- \Delta J}{2}\right) \right]}^N.
\label{eqn-Zc}
\end{equation}

The corresponding expression for the average energy is, 
 \begin{equation}
\langle E \rangle = \frac{N \Delta J}{2}\left[\frac{1}{ 1 +  \exp( \Delta J/2)} \right ]\equiv N\epsilon
\label{eqn-E}
\end{equation}
\\
Fig.~\ref{fig2}(a) displays the average energy per particle $\epsilon$  vs $\Delta J$  from analytical expression of Eq.~\ref{eqn-E} and its comparison with the simulation results. 

While the average polarization is zero, the correlation function for the polarization as a function of  particle distance, $r$,  assumes the form, 
\begin{equation}
 G(r) ={\left( \frac{1 - \exp(-\Delta J/2)}{1 + \exp(-\Delta J/2)}\right )}^{r}
\end{equation}
The corresponding expression for the correlation length is, 

\begin{equation}
    \xi = {| \ln(1 - e^{-\Delta J/2}) -\ln(1 + e^{-\Delta J/2})|}^{-1}
\end{equation}
In the limit of $\exp(\Delta J/2) >> 1$, the correlation length $\xi \rightarrow \frac{1}{2}\exp(\Delta J /2)$. Thus as would be expected for equilibrium 1D systems, there is no long range correlation of the polarization. However, for any finite lattice size system of size $L$, as long as $\xi > L$, the confluent state would tend to exhibit a polarized state.

In the presence of an external field $h$ which couples with the individual particle polarization, the corresponding  Hamiltonian   assumes the   form,
\begin{equation}
H = \sum_{i} J_1 \Theta( \sigma_{i} - \sigma_{i+1} ) -J_2 \Theta( \sigma_{i+1} - \sigma_{i} )  - h \sigma_{i} \nonumber
\label{eq2}
\end{equation}

\begin{figure*}[t]
 \centering
    \includegraphics[width= \linewidth, height = 8cm]{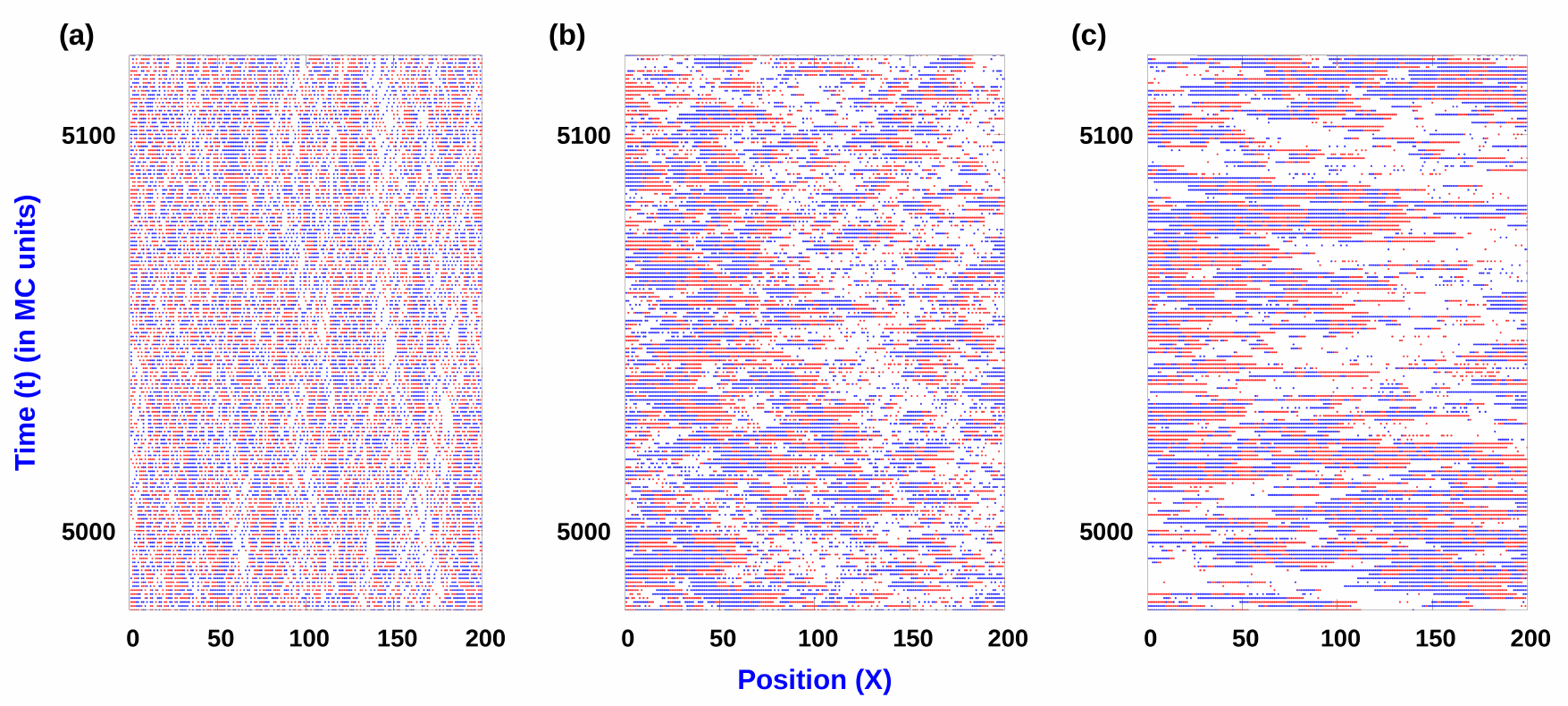}
    \caption{Spatio-temporal plot: Time snapshots of distribution of right polarized $(+)$(blue) and left polarized $(-)$(red) particles on the lattice. Here (a)$Q = 0.1$, (b) $Q = 10$, $Q = 50$, $J_1 = 4$, $J_2 = 0$, with $\rho = 0.6$. MC simulations where done with $L = 1000$} 
    \label{fig3}
\end{figure*}  

The corresponding expression for the Gibb's Canonical Partition Function $Z_g$ is, 

\begin{eqnarray}
Z_g &=& \sum_{[\sigma_i]} \exp \left[ - \sum_{i} J_1 \Theta( \sigma_{i} - \sigma_{i+1} ) -J_2 \Theta( \sigma_{i+1} - \sigma_{i} ) + h \sum_{i} \sigma_i \right], \nonumber
\end{eqnarray}
\noindent and the corresponding transfer matrix reads
\[
T=
  \begin{bmatrix}
    e^{h} & e^{- J_{1}} \\\\
    ~e^{J_2} & e^{- h}
  \end{bmatrix}
\]

Again, using transfer matrix method, we obtain 
\begin{equation}
Z_g = {\left[ \cosh(h) + [ {\cosh^2 (h) + e^{ -\Delta J} -1 ]}^{1/2} \right]}^{N}
\label{eqn-Zg}
\end{equation}
Using this, the expression for the average polarization of the particle, $\sigma$ is,

\begin{equation}
\sigma = \frac{\sinh(h)}{\left[ \cosh^{2}(h) + e^{ -\Delta J} -1\right]^{1/2}}
\label{eqn-m}
\end{equation}
Fig.\ref{fig2}(b)  shows the excellent agreement  of the variation of the average polarization with external field $h$ obtained from  Eq.~\ref{eqn-m} and with the simulation. 

\begin{figure*}[t]
 \centering
    \includegraphics[width = \linewidth, height = 7cm]{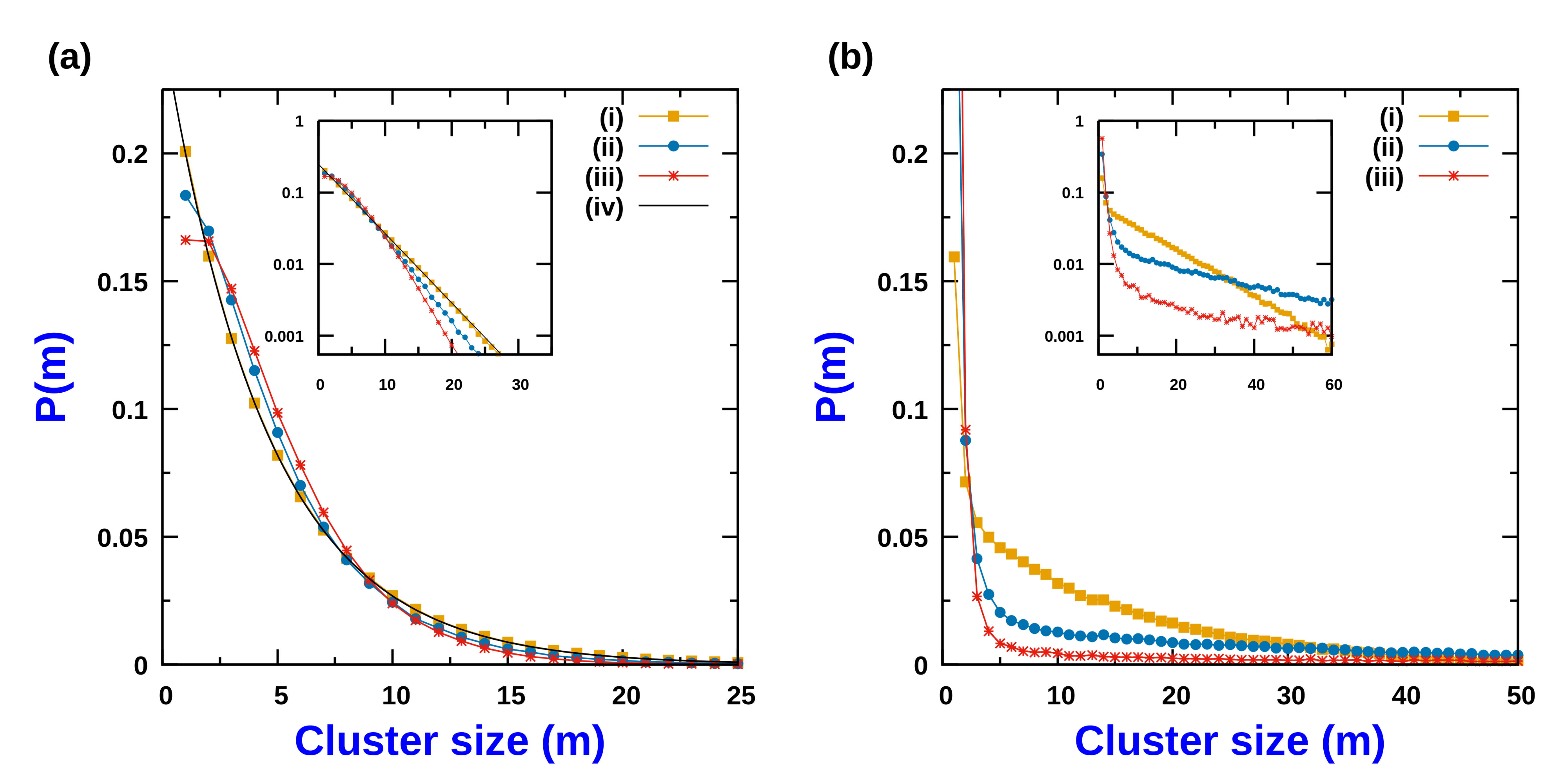}
    \caption{(a) Cluster size (m) probability distribution in the low $Q$ regime;  $Q = 0.1$  for different CIL strength: (i) $J_1 = 0.1$, (ii) $J_1 = 3$, (iii)$J_1 = 7$, (iv) Eq.\ref{eqn-SEP} ( Probability density function for TASEP and SEP). (b) Cluster size (m) probability distribution in the  high $Q$ regime; $Q = 30$  for different CIL strength: (i) $J_1 = 0.1$, (ii) $J_1 = 3$, (iii)$J_1 = 7$. The inset figures are the  corresponding logplots.  In all cases, $J_2 =0$, $\rho = 0.8$ and $L=1000$. MC simulations are performed and averaging is done over 2000 samples.}     
    \label{fig4}
\end{figure*}  

\begin{figure}[t]
 \centering
    \includegraphics[width= \linewidth, height = 7cm]{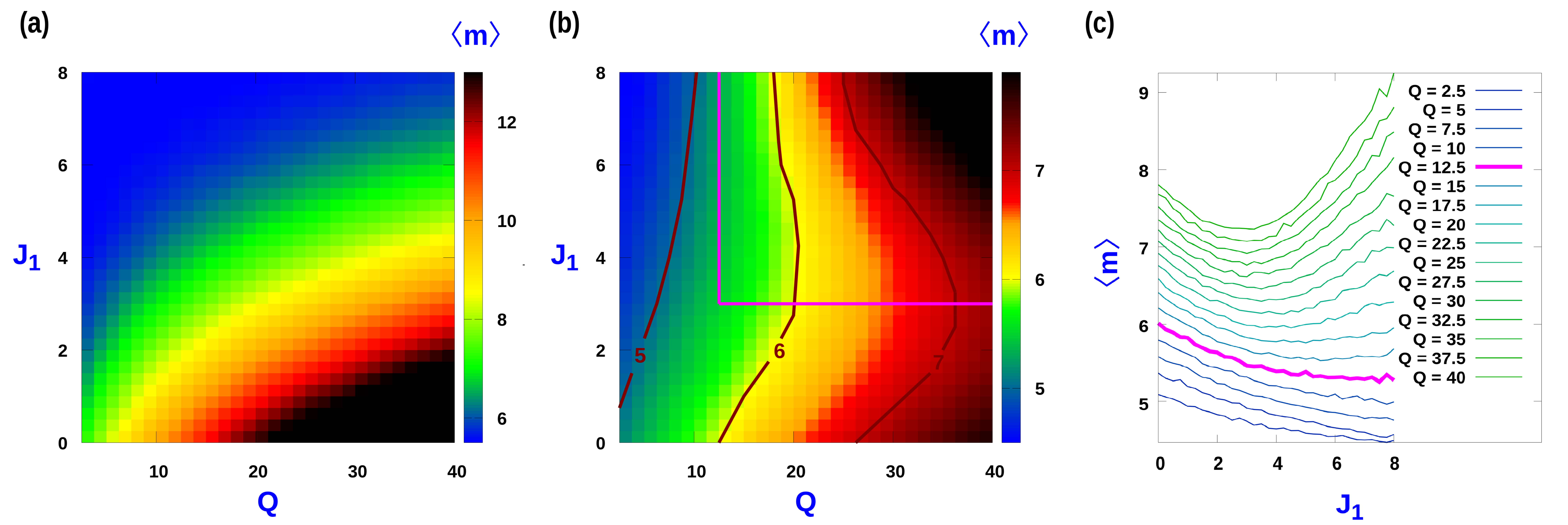}
    \caption{(a) Contour plot of average cluster size $\langle m \rangle $ as function of $J_1$ and $Q$ when $J_1 = J_2$ i.e. $\Delta J = 0$. (b) Contour plot of average cluster size $\langle m \rangle $ as function of $J_1$  and $Q$ for  constant $J_2 (J_2 = 4)$: Re-entrant like behaviour is observed for $Q > Q_c$ with $Q_c = 12.5$. (c)  Plot of $\langle m \rangle $ with $J_1 $ corresponding to (b). For all cases, MC simulations where done with $L = 1000$ and averaging was done over 2500 samples at $\rho = 0.8$.}
    \label{fig5new}
\end{figure}  

\subsection{ Clustering in the presence of vacancies}

We next focus our attention on the nature of clustering behaviour of the particles that arises out of the interplay of translation dynamics of particles and the switching dynamics of particles.

The CIL interaction is controlled by the parameters $J_1$ and $J_2$. While $J_1$ is the energy cost associated with the polarities of neighbouring particles pointing towards each other, $J_2$ is the reduction of energy associated with the polarities of the neighbouring particles pointing away from each other. In the absence of CIL  between particles in a cluster, the switching rate from $(\rightarrow)$ to $(\leftarrow)$ is $b$ and it is identical to the switching rate between $(\leftarrow)$ to $(\rightarrow)$. We define a dimensionless quantity $Q$, which is the ratio of the translation rate and which behaves as a measure of cell activity, $a$ and the switching rate, $b$, and set $b = 1$ without loss of generality. In general, clustering will be controlled by the strength of CIL and $Q$, although the overall vacancy density is a conserved quantity, and it will also affect the cluster size distribution. In Fig. \ref{fig3}, we display the spatio-temporal evolution of the clusters for different activity strength quantified in terms of $Q$, for a fixed value of CIL interaction strength $J_1$ and holding $J_2 = 0$. At relatively high values of $Q$ ( See Fig.\ref{fig3}b and Fig.\ref{fig3}c), the system segregates into alternating domains of dense clusters  and a low density  gas region. The mean sizes of these dense clusters increases on increasing $Q$. For the  dense clusters, the dynamics of their sizes is governed by the processes at the cluster boundaries. Typically, for large $Q$ regime, the composition of the dense cluster is such that an array of right pointing $(\rightarrow)$ particles occupy the left end of dense cluster domain while the right end comprises of an array of left pointing $(\leftarrow)$ particles. Thus the internal structure within the bulk of such dense cluster comprises of  defects- pairs of $(\rightarrow)$ and $(\leftarrow)$ particles. It may also be noted that such dense cluster for which the polarity of the particles at both ends  are pointing inwards are immobile. At any instant of time, a cluster size can increase if a $(\rightarrow)$ particle from the adjoining gas region joins the left end of the dense cluster or if a $(\leftarrow)$ particle from the adjoining gas region joins the right end of the dense cluster. On the other hand, a cluster size can decrease if a single particle at the cluster ends switches it's polarity and subsequently  leaves the cluster.

We next investigate the effect of CIL and strength of  activity on the cluster size distribution. First we choose  $J_2 = 0$, and modulate the CIL  strength by variation in $J_1$. Fig.\ref{fig4} displays the cluster size distribution for different CIL interaction strength ($J_1$) and strength of activity $Q$. The limit - $Q << 1$, corresponds to a situation of  low cell activity when  particle translation rate is much slower than its polarity switching rate. For this case, in the absence of CIL interaction $( J_1 = J_2 = 0)$,  the probability of m-particle cluster is simply proportional to $\rho^m(1- \rho)$ for $N \rightarrow \infty$, where, $\rho$ is the number density of particles on the lattice  \cite{pulki}. Consequently, the normalized probability of m-particle cluster is ,  $P(m) = \rho^{m-1}(1-\rho)$ and which maybe expressed in the exponential form, 

\begin{equation}
P(m) = \left(\frac{1-\rho}{\rho}\right)e^{-m/\xi},
\label{eqn-SEP}
\end{equation}

\noindent where $\xi = {| ~{\ln \rho} ~|}^{-1}$, and  the mean  cluster size reads $\langle m \rangle = {(1-\rho)}^{-1}$. Eq.~\ref{eqn-SEP} is identical to the cluster size distribution of a Totally asymmetric exclusion process (TASEP) and Symmetric exclusion process (SEP)~\cite{pulki}.  For low activity,  increasing the CIL strength does not change the exponential nature of the cluster size distribution (See Fig.~\ref{fig4}a). Although there is significant deviation from the exponential decay  for small size clusters at large activity levels, $Q >> 1$, the asymptotic decay remains exponential as indicated by the log-plot of the distribution  in Fig.~\ref{fig4}(b),  irrespective of the strength of CIL. As shown in Fig.~\ref{fig4}(a) and Fig.~\ref{fig4}(b), increasing the activity $(Q)$  leads to an increase in  the average cluster size, and to a broadening of the cluster size probability distribution both in the presence and absence of CIL. This generic trend comes from the accumulation of active particles in regions where they move slowly~\cite{schnitzer,active-rev}. Due to excluded volume interaction, motility vanishes when a particle is obstructed by a neighbouring particle leading to formation of a cluster. The restoration of the mobility in a cluster is related to the cluster disintegration, which in turn  depends on the switching rate of the particle's polarity  at the cluster edges. Consequently,  lowering the switching rate (hence increasing $Q$) increases  the average cluster size. 

To unravel the role of CIL on the average cluster size, we scrutinize separately the impact of $J_1$ and $J_2$. Within a cluster, while $J_1$ favours polarity alignment, $J_2$ favours polarity anti-alignment of neighbouring particles, and accordingly, the interaction term of the Hamiltonian associated with CIL breaks the symmetry associated with independent rotation of the polarity vector. When
$J_1 = J_2$, i.e., $\Delta J = 0$, the CIL term in the Hamiltonian is identical to the CIL interaction considered on a phenomenological hydrodynamic theory for 1D cell assemblies \cite{ananyo}. In this regime, increasing CIL disfavours large cluster formation, leading to a monotonic decrease of the average cluster size with increasing CIL irrespective of the activity strength, $Q$. In this limit, the overall behaviour of the cluster size as a function of $Q$ and $J_1$ maybe quantified in terms of the contour plot displayed in Fig.\ref{fig5new}(a). If $J_1 \neq J_2$, which corresponds to an asymmetric CIL interaction, the effect of $J_1$ on the cluster size is more involved. Increasing $J_1$ at fixed $J_2$ has two effects on the clusters. While $J_2$ enhances the tendency of particles at the cluster edge to flip outwards, $J_1$ favours a polar alignment among neighbouring particles. This coupling favours overall that particles point inwards. Hence, $J_2$ promotes decrease in cluster size  while $J_1$ has the opposite effect. Therefore, we can expect larger clusters to be stabilized when $J1 > J2$. In general, the competition between the opposite effects associated to $J_1$ and $J_2$ lead to a non-monotonic behaviour of the average cluster size, $\langle m \rangle $, as a function of $J_1$ beyond a threshold value of activity $Q$.  Fig.\ref{fig5new}(b), which displays the contour map of $\langle m \rangle$, for a fixed value of   $J_2 = 4$, illustrates how the re-entrant like behaviour features for a wide range of $Q$ and $J_1$. The corresponding non-monotonic behaviour of $\langle m \rangle$ as a function of $J_1$ for different sets of $Q$, is displayed in Fig.\ref{fig5new}(c).

\subsection{Approximate expression  for average cluster size }

When the particle translation rate   is much faster than the polarity switching  rate, $(Q >> 1)$, the system segregates into alternate regions of dense clusters (c)  and a low density gas phase (g). In this regime, the  interaction between the dense clusters is weak, and the stationary state is achieved as a  balance between the  incoming particle  flux  into the dense clusters from the surrounding gas region and the outgoing particle  flux  from the boundaries of the dense cluster region due to the particle switching their polarity at the  cluster boundaries ~\cite{soto}. This balance can be estimated from an equivalent equilibrium process  where the size distribution of dense clusters is determined by minimizing the effective Helmholtz free energy. It is worthwhile to point out that in the absence of CIL, the corresponding cluster size distribution  is obtained by minimizing the system configurational entropy,  as has been  done for persistent exclusion process (PEP)~\cite{soto}. 

\subsubsection{ \bf Minimization of effective Helmholtz Free energy (F)}

Without loss of generality, for simplicity we incorporate the effect of CIL through $J_1$, and set $J_2 =0$. However,  the procedure outlined to obtain the form of cluster size distribution can easily be generalized to $J_2 \neq 0$. From Eq.~\ref{eqn-E} we can express the mean energy of a cluster of length $l$ as
\begin{equation}
\langle E(l) \rangle = \frac{l}{2}\left[\frac{J_1}{ 1 +  e^{J_1/2}} \right ].
\end{equation}
\noindent The configurational entropy $S$ of the dense cluster phase corresponds to the number of ways by which $\Omega$  clusters  can be arranged such that the clusters of same length are indistinguishable and are subject to the constraint that the total number of sites occupied by the clusters, $N_c$, is  constant. Then by fixing   $\Omega$, the entropy can be expressed as,

\begin{eqnarray}
S = \ln\left[ \frac{\Omega ~!}{ \displaystyle\prod_l G_c(l)!} \right] - \lambda \left ( N_c - \sum_{l} l G_c(l) \right ) \nonumber  - \gamma \left ( \Omega - \sum_{l} G_c(l) \right ),
\end{eqnarray}
\noindent
where $G_c(l)$  is the number of clusters of length $l$ in the cluster (c) phase and $\lambda$ and $\gamma$ are Lagrange multipliers, as already proposed for a PEP process~\cite{soto}. 
Accordingly, the free energy, $F$, of the dense phase can be derived by  accounting for  the previous configurational entropy and the cluster internal energy due to CIL, leading to
\begin{eqnarray}
F = \left[\frac{J_1 /2 }{ 1 +  e^{J_1/2}} \right] \sum_{l} l G_c(l) - \ln \left[ \frac{\Omega ~!}{ \displaystyle\prod_l G_c(l)!} \right]  + \lambda \left ( N_c - \sum_{l} l G_c(l) \right ) +  \gamma \left ( \Omega - \sum_{l} G_c(l) \right ) \nonumber
\end{eqnarray}
The corresponding expression for the variation of the free energy, $\delta F$, reads 
\begin{equation}
\delta F =  \sum_{l} \left[~\ln G_c(l)  + \left( \lambda  + \frac{J_1 /2 }{ 1 +  e^{J_1/2}} \right)l + \gamma ~ \right] \delta G_c(l)
\end{equation}
and its minimization, $\delta F = 0$ for independent variations of $\delta G_c(l)$, provides cluster size distribution
\begin{equation}
G_c(l) = A_c e ^{-l/l_c},
\end{equation}
which has an exponential shape.\\
A similar argument for the gas(g) phase yields $G_g(l)$ -the number of clusters of length $l$ in the gas(g) phase, with the corresponding form being, 
\begin{equation}
G_g(l) = A_g e ^{-l/l_g}
\end{equation}

The corresponding parameters ,  $A_g$, $A_c$, $l_g$ and $l_c$, that characterize  univocally the  coexisting cluster size distributions of the gas and dense phase can be determined  imposing

\noindent
(a) The total number of clusters in the gas phase must equal total number of clusters in the dense cluster phase. This implies,
\begin{equation}
\sum_{l} G_c(l) = \sum_{l} G_g(l) = \Omega
\label{cond1}
\end{equation}
(b) The total number of sites occupied by clusters in the gas phase together with the total number sites occupied by clusters in the dense cluster phase  must equal total number of lattice sites,
\begin{equation}
\sum_{l} l G_c(l) + \sum_{l} l G_g(l) = N 
\label{cond2}
\end{equation}
(c) If $\phi_c$ and  $\phi_g$ are  the number density of particles in dense cluster region and gas region, with  $\langle l_c \rangle$ and  $\langle l_g \rangle$ being  the average length of the dense cluster and gas region, then the overall particle density $\rho$ should obey,
\begin{equation}
\langle l_c \rangle \phi_c + \langle l_g \rangle \phi_g = \left[~\langle l_c \rangle + \langle l_g \rangle ~\right]~ \rho
\label{cond3}
\end{equation}

\noindent
(d) If the hopping rate is much larger than switching rate, the gas region has typically a very low density of particles , hence   $\phi_g<<1$. When a switching event at the boundary of a dense cluster occurs, a particle is emitted into the gas region. This  leads to production of a dimer within the gas region which are the dominant clusters in the gas phase.  Accordingly, we invoke the condition  that in the limit of  $Q >> 1$, {\it the steady state is determined by the condition of matching the dimer  production and disintegration rates}~\cite{soto}. 

The dimer production   in the gas  occurs when a particle at the edge of the adjoining dense cluster (with its polarity pointing inwards towards the bulk of cluster), has flipped its polarity and thus breaks away into the gas region, and within the time $\tau$ that it takes to reach the adjacent dense cluster, the particle at the edge of the adjacent dense cluster region has flipped its polarity.  The average time that it takes for a particle in edge of the dense cluster to reach the edge of the adjacent dense cluster reads  $\langle \tau \rangle=\langle l_g \rangle / a$. Assuming that the neighbours of the edge particle of the dense cluster are pointing inwards, the rate of flipping of the particle at the edge is approximately equal to $b$.
Hence, the overall dimer production rate, $W_2^{p}$, may be approximated by,
\begin{equation}
W_2^{p} = 2 b^{2}\frac{ \langle l_g \rangle}{a} \sum_{l} G_g(l)
\label{dimer1}
\end{equation} 
\noindent where the contribution  from trimer disintegration is  neglected. 

A typical configuration of a dimer would be $(\rightarrow)(\leftarrow)$. The disintegration of a dimer would occur when {\it either} of the particles of the 2 particle dimer cluster switches their direction of polarity. While in the absence  of CIL  the switching rate is $b$, in the presence of CIL the switching rate is $b e^{J_1/2}$. 
Therefore
\begin{equation}
W_2^{c} = 2 b e^{J_1/2} ~G_c(2)
\label{dimer2}
\end{equation}
In the  steady state we equate Eq.~\ref{dimer1} and Eq.~\ref{dimer2} to obtain the condition

\begin{equation}
Qe^{J_1/2} G_c(2) =  \langle l_g \rangle \sum_{l} G_g(l).
\label{cond4}
\end{equation}
\noindent Since  the particle flux  from the gas into the cluster region, $a \phi_g/2$, must equal the particle flux  from the dense cluster region,  $b$ (due to  particle switching at the cluster boundary), at steady state, we arrive at

\begin{equation}
\phi_g = 2/Q
\end{equation}
 
\begin{figure}[t]
 \centering
    \includegraphics[width= \linewidth, height= 7cm ]{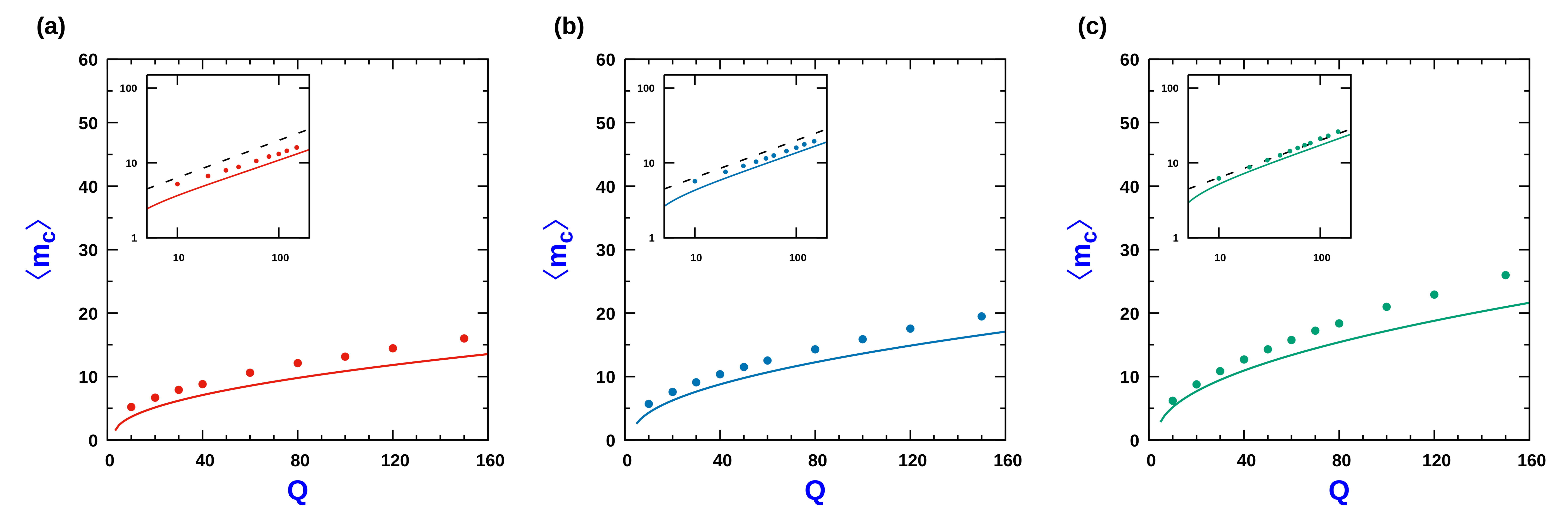}
   \caption{Variation of the average cluster size, $\langle m_c \rangle$, in the cluster phase as a function of $Q$: Comparison of the theoretical expression, Eq.~(\ref{MF-cluster}) ,with MC simulation. (a) $J_1 = 0$ (No CIL). (b) $J_1 = 1$ and (c) $J_1 = 2$. In all cases, $J_2 = 0$ and $ \rho = 0.8$. Inset figures are the corresponding log-log plots. Solid lines correspond to expression in Eq.~(\ref{MF-cluster}) while the circles correspond to MC simulations for each case. The dashed lines in the inset log-log plot corresponds to a slope of $1/2$. MC simulation done with $L = 2000$ and averaging was done over 3000 samples.}
    \label{fig6}
\end{figure}  

\subsubsection{\bf Mean cluster size}

We approximate the sums in Eq.(\ref{cond1}),  Eq.(\ref{cond2}) and  Eq.(\ref{cond4}) by integrals to obtain an approximate expression for the mean cluster size. We set the integration limit for the dense cluster phase between $2$ and $\infty$ since, for having a dense cluster, there needs to be at least 2 particles. For the gas phase, there has to be at least 1 site, which sets the lower integration limit. The upper integration limit is taken to $\infty$ for $N \rightarrow \infty$. With these conversions, we have,
\begin{eqnarray}
\sum_{l} G_c(l) &=& A_c l_c e^{-2/l_c}\nonumber \\
\sum_{l} G_g(l) &=& A_g l_g e^{-1/l_g}\nonumber \\
\sum_{l} l G_c(l) &=&A_c l_c e^{-2/l_c} (2 + l_c) \nonumber \\
\sum_{l} l G_g(l) &= & A_g l_g e^{-1/l_g} (1 + l_g) \nonumber 
\end{eqnarray}
Substituting these expressions in Eq.(\ref{cond1}-\ref{cond4}), we obtain the algebraic relations involving $A_c$, $A_g$, $l_g$ and $l_c$, 
\begin{eqnarray}
A_c l_c e^{-2/l_c}  &=& A_g l_g e^{-1/l_g} \\
(2 + l_c)A_c l_c  e^{-2/l_c} + (1 + l_g) A_g l_g e^{-1/l_g} &=& N  \\
(2 + l_c) + \frac{2}{Q}(1 + l_g) &=&  (2 + l_c + 1 + l_g)\rho   \\
Qe^{J_1/2} A_c e^{-2/l_c} &=& (1 + l_g)A_g l_g e^{-1/l_g}
\end{eqnarray}
The previous approximation leads to, 
\begin{eqnarray}
\langle m_g \rangle &=&  1 + l_g\nonumber\\
\langle m_c \rangle &=&  2 + l_c 
 \label{mc}
 \end{eqnarray} 
Using these relations, we arrive at the mean cluster size prediction 

\begin{equation}
\langle m_c \rangle  = 1 +  \sqrt{ 1 + \left( \frac{\rho Q - 2}{1- \rho}\right) e^{J_1/2} }
\label{MF-cluster}
\end{equation}
\noindent  
Fig.~\ref{fig6} compares  the analytic prediction of  $\langle m_c \rangle$ with the simulation results for a range of CIL strengths.
While in the absence of CIL (Fig.\ref{fig6}a), or weak CIL strength (Fig.\ref{fig6}b), the approximate analytical form matches the MC simulation results reasonably well, for moderate CIL strength $(J_1 =2 )$, the deviation from the simulation result is more pronounced although even for this case $\langle m_c \rangle \sim Q^{1/2}$ when $Q >> 1$ (See Fig,\ref{fig6}c). However for very strong CIL interaction strength, the approximate analytical curve fails to reproduce average cluster size characteristics. 

\begin{figure*}[t]
 \centering
    \includegraphics[height = 7 cm, width = \linewidth]{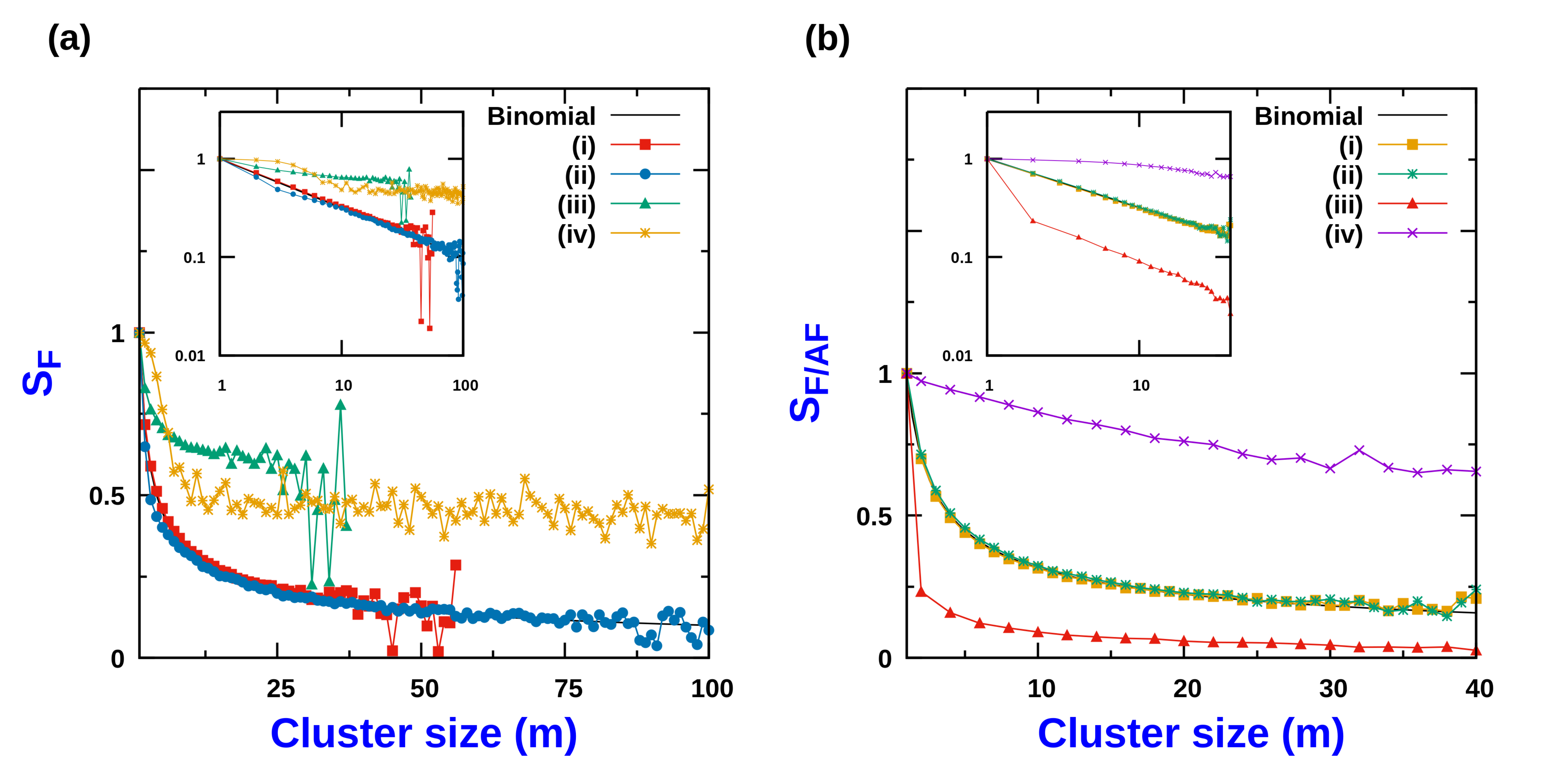}
    \caption{ (a) RMS Fluctuation of polarization, $(S_F)$ with cluster size $(m)$: (i) No CIL and low Q; $Q = 0.1$, (ii) No CIL and high Q; $ Q = 30$, (iii) $J_1 = 7$ and low Q; $Q = 0.1$, (iv) $J_1 = 7$, high Q; $Q = 30$. Here. $J_2=0$. The binomial distribution corresponds to solid black line. (b) RMS Fluctuation of local antiferromagnetic order parameter,$ S_{AF}$, and  ferromagnetic order parameter, $ S_F$, in a cluster of size m  vs Cluster size (m) : (i) $S_F$ for No CIL and low Q; $Q = 0.1$, (ii) $S_{AF}$ for No CIL and low Q; $Q = 0.1$, (iii) $S_{F}$ for $J_2 = 7$ and high Q; $Q = 30$, (iv) $S_{AF}$ for $J_2 = 7$, high Q; $Q = 30$. Here, $J_1 = 0$. MC simulations were done with $L = 1000$ and averaging over 2000 samples.} 
    \label{fig7}
\end{figure*}  

\subsection{Cluster Polarisation}
CIL has a strong impact not only on host particle aggregate in clusters, but also on how they  align relative to each other. The net  polarization, of  a cluster of size $m$,  is quantified by  $P_m =  \sum_{i =1}^{m}  \sigma_i $. However, since clusters generically do not  develop a net polarity, we analyze  their root-mean square (RMS) fluctuations, 
\begin{equation}
 S_F = \frac{1}{m} \sqrt { \overline{ {\left( \sum_{i =1}^{m}  \sigma_i  \right)}^{2} } } 
 \end{equation}
In the absence of CIL when $Q << 1$, particle's polarization are uncorrelated from  the neighbouring ones. In this regime, as displayed in Fig.\ref{fig7}a(i),  $ S_F = m^{-1/2}$, corresponding to the RMS of a binomial distribution. For large activity, $Q >> 1$, in the absence of CIL, for relatively small cluster sizes, there is a negative deviation of $S_F$ compared to  $ m^{-1/2}$, although for sufficiently large clusters, it merges with the form corresponding to binomial distribution as displayed in Fig.~\ref{fig7}a(ii). In the presence of CIL  for which $J_1$ is high, and $J_2 = 0$, for range of cluster sizes, $S_F$ does not scale as $m^{-1/2}$ and instead remains virtually unchanged irrespective of whether activity is high or low as displayed in Fig.\ref{fig7}a(iii) and \ref{fig7}a(iv)) respectively. However the cases of high and low activity are distinguished by the fact that the formation of large size clusters is much more prevalent when activity is high as compared to the case when activity is low. 
  
When $J_2 \neq 0$ and $J_1 =0$,  CIL induces "anti-ferromagnetic" like order within the cluster. This ordering is captured through
the  "anti-ferromagnetic"-like order parameter defined in terms of  fluctuation of the difference of polarization of two sublattices, $A$ and $B$ within a cluster of size $m$,
\begin{equation}
 S_{AF} = \frac{1}{m} \sqrt { \overline{ {\left( \sum_{i =1}^{m}   (\sigma_{i}^{A} - \sigma_{i}^{B} )  \right)}^{2} } } 
 \end{equation}
\noindent where the summation over $i$ for $A$ sublattice is over odd sites and for $B$ sublattice it is over the even sites within a cluster. In the absence of CIL interaction, when $Q << 1$, the polarization of the individual particles are uncorrelated with the polarization of the other particles constituting the cluster. Hence in this limit, both the ferromagnetic and anti-ferromagnetic order parameter, e.g., $S_F$ and $S_{AF}$ are identical to RMS fluctuation of a binomial distribution with    $S_F = m^{-1/2}$ as displayed in Fig.~\ref{fig7}b(i) and \ref{fig7}b(ii).
On the other hand when $Q>>1$, when $J_2$ is high (and $J_1 = 0$), the anti-ferromagnetic order parameter, $S_{AF}$ has a positive deviation from $m^{1/2}$ for all ranges of cluster size, while the ferromagnetic OP ($S_F$) has a negative deviation as shown in Fig.\ref{fig7}b(iii) and \ref{fig7}b(iv).

\begin{figure*}[t]
 \centering
    \includegraphics[height = 7 cm, width = \linewidth]{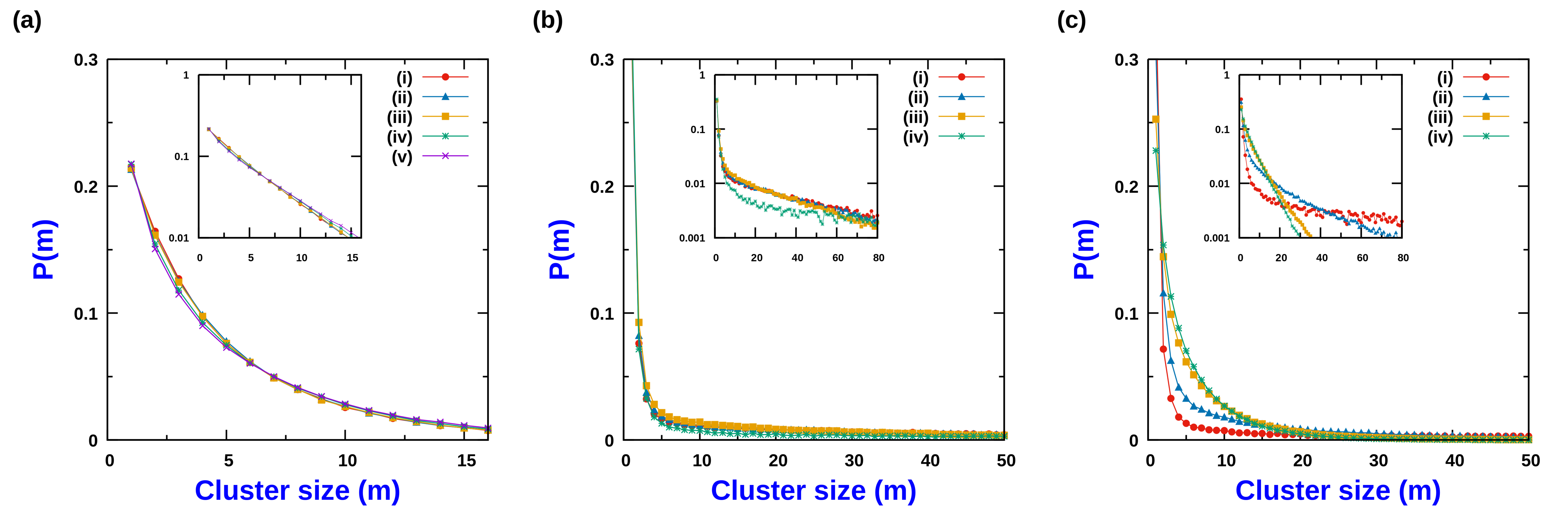}
    \caption{Cluster size distribution function $P(m)$ : 
    (a) when $h_o \leq a$ for low $Q$ ( $Q = 1$) with $a =1$: (i) $h_o = 0$, (ii) $h_o = 0.2$, (iii) $h_o = 0.5$, (iv) $h_o = 0.9$, (v) $h_o = 1$. (b) when $h_o \leq a$ for high $Q$ ( $Q = 50$) with $a = 50$: (i) $h_o = 0$, (ii) $h_o = 25$, (iii) $h_o = 40$, (iv) $h_o = 50$. (c) when $h_o \geq a$ for high $Q$ ( $Q = 50$) with $a = 50$: (i) $h_o = 50$, (ii) $h_o = 52$, (iii) $h_o = 75$, (iv) $h_o = 100$. Inset figures are corresponding log-log plots. For all cases, $J_1 = 3$, $J_2 = 0$ and $\rho = 0.8$. MC simulations were done with $L = 1000$ and averaging was done over 2000 samples.} 
    \label{fig8}
\end{figure*} 

\subsection{Effect of external field on cluster characteristics }

We incorporate  the effect of an external field $h_o$ whose effect is to aid translation of right end directed $(\rightarrow)$  particles and oppose motion of left end directed $(\leftarrow)$ particles. Effectively the hopping rate of the  $(\rightarrow)$ particle becomes $a + h_o$, while  for $(\leftarrow)$ particles the hopping rate is $a-h_o$. As long as $h_o < a$, the natural direction of movement of both $(\rightarrow)$ and $(\leftarrow)$ particles is retained. When $h_o = a$, the $-$ ended particles  do not move. For $h_o > a$, both $(\rightarrow)$ and $(\leftarrow)$ particles move in the same direction, as prescribed by the external field. For   relatively low activity, when $h_o < a$,  increasing  $h_o$ has the effect of marginally increasing the average cluster size. In general, the cluster size is not significantly altered ( See  Fig.\ref{fig8}a). 

 As shown in Fig.~\ref{fig8}(b), for large activity the average cluster size varies non monotonically with $h_o$. Initially, the average cluster size decreases, but after a threshold, the mean cluster size starts increasing with the external field magnitude. However for sufficiently high strength of external field $h_o$, the  average cluster size again starts increasing and the corresponding distribution function of the cluster size starts broadening with $h_o$. As $h_o$ approaches the hopping rate $a$, average cluster size sharply increases. Finally when $h_o = a$, for which the $(\leftarrow)$ particles stop moving, the cluster size distribution becomes broadest, with average cluster size increasing drastically (around 2.5 times the average cluster size for $h_o = 0$). As shown in Fig.~\ref{fig8}(c), when $h_o$ is increased further, $h_o > a$, not only do  $(\rightarrow)$ particles and $(\leftarrow)$ particles move in the same direction, the corresponding average cluster size monotonically decreases with increase in $h_o$.

\section*{Discussion}

In this paper we have presented a minimal discrete driven  lattice gas model which  mimics the phenomenology of  CIL interactions between cells and the movement of the cells in confined in 1D channel.  In the absence of vacancies ( akin to dense packing of cells in 1D array), the translation dynamics is arrested. In this limit, the model reduces to an equilibrium spin model which does not possess $Z_2$ symmetry like Ising model. We solve this  model exactly both in the presence and absence of an external field which couples to the individual polarization of the cell. 
In the presence of vacancies, the interplay of translation dynamics and CIL interaction between particles results in  a steady size cluster size distribution which is exponentially distributed for the large size cluster. The typical cluster size and the distribution function is controlled by CIL strength and activity. The effect of increasing activity at fixed CIL strength invariably leads to an increase in the average cluster size. However the effect of varying  CIL interaction ( through $J_1$ at constant $J_2$), can result in a non-monotonic dependence of the average cluster size as a function of CIL strength $J_1$. In the high activity regime,  $Q >> 1$, an analytic form of  average cluster size can be obtained approximately by effectively mapping the system to an equivalent equilibrium process involving of clusters of  sizes  wherein the cluster size distribution is obtained by minimizing an effective Helmholtz free energy. The resultant prediction of exponential dependence on $J_1$ of the average cluster size and $Q^{1/2}$ dependence of the average cluster size is borne out to reasonable accuracy as long as $J_1$ is not very large. At very high values of $J_1$, the prediction breaks down, indicating the failure of this approximation scheme. The polarization characteristics within an cluster is indicated by emergence of a local ferromagnetic like order parameter within a cluster due to the effect of CIL, when $J_1  > J_2$. In the presence of an external field $h_o$ which couples to translation dynamics of individual system, system, such that  it aids translation of right end directed $(\rightarrow)$ particles and opposes motion of left end directed $(\leftarrow)$ particles, we find that the the average cluster size exhibits a non-monotonic dependence on $h_o$. While many studies of active particles systems have predicted and reported existence of phase transitions due to Motility induced Phase separation(MIPS) \cite{active-prl,active-rev}, we have not observed any such phase transitions for our model for the range of parameters that we have explored.

While in  our present work we have focused solely on the interplay of  cell movement and CIL interaction between cells, and as such treated  the cells as hard objects, its worthwhile to point out that  apart from re-polarization events,  adhesion of the colliding cells and  the relatively rare event of cells walking past each other has also been observed \cite{cil1,cil2}. Cells are contractile entities which exert forces on each other and on the underlying substrate. It has been observed that cells on contact with each other make use of  trans-membrane adhesion molecules such as  cadherins to adhere to each other \cite{cadherin}. Presence of these adhesion molecules serve to generate an effective attractive force between the cells. Thus a more holistic understanding of the collective organization of cells would require inclusion of attractive interaction between the cells, additional alignment mechanisms that might be at play apart from the effect of CIL that we have discussed. In our future work we seek to incorporate these aspects in the framework of our modeling approach. Finally we also seek to generalize our discrete lattice gas modeling approach to study the effect of CIL in context cellular organization on 2D substrate.  

\bibliographystyle{prsty}



\section*{Acknowledgements}

Financial support is acknowledged by H.P. for Mahajyoti fellowship program (MJRF-2022). S.M. acknowledges financial support and hospitality for visit to ICTP, Trieste under the Associateship program, where part of the work was done. S.M. acknowledges useful discussion with Amitabha Nandi (IIT Bombay) and Deepak Dhar (IISER, Pune). 

\section*{Author contributions statement}

H.P. carried out simulations, analyzed model and data, I.P. designed study, analyzed model and data and wrote manuscript. S.M. designed study, analyzed model and data, formulated and carried out theoretical calculations and wrote the manuscript. All authors reviewed the manuscript. 

\section*{Competing interests}

Authors declare no competing interests.






\end{document}